\documentclass[twocolumn, showpacs, prl]{revtex4}
\usepackage{graphicx}
\usepackage{dcolumn}
\usepackage{amsmath}
\begin{document}
\hsize\textwidth\columnwidth\hsize\csname@twocolumnfalse\endcsname
\bibliographystyle{prl}
\title{Density dependent spin polarisation in ultra low-disorder quantum wires}
\author{D. J. Reilly,$^{* 1,2}$ T.  M.  Buehler,$^{1,2}$ J. L. O'Brien,$^{1,2}$ A.  R.  Hamilton,$^{1,2}$ 
A.  S.  Dzurak$^{1,3}$, R.  G.  Clark,$^{1,2}$ B.  E.  Kane$^{\dagger}$,  
L. N. Pfeiffer$^4$ and K. W. West$^4$}
\affiliation{$^1$Centre for Quantum Computer Technology, University of New
South Wales, Sydney 2052, Australia}
\affiliation{$^2$School of Physics, University of New South Wales, Sydney 2052, Australia} 
\affiliation{$^3$School of Electrical Engineering \& Telecommunications, 
University of New South Wales, Sydney 2052, Australia}
\affiliation{$^4$Bell Laboratories, Lucent Technologies, Murray Hill, New
Jersey 07974}

\begin{abstract}
\noindent
There is controversy as to whether a one-dimensional (1D) electron gas can spin 
polarise in the absence of a magnetic field.  Together with a simple 
model, we present conductance measurements on ultra low-disorder quantum wires 
supportive of a spin polarisation at $B=0$. A spin energy gap is indicated by 
the presence of a feature in the range $0.5 - 0.7 \times 2e^2/h$ in conductance data. 
Importantly, it appears that the spin gap is not static but a function of 
the electron density.  Data obtained using a bias spectroscopy technique 
are consistent with the spin gap widening further as the Fermi-level is increased.  
\end{abstract}

\maketitle
In the presence of strong exchange coupling, electrons can spin polarise in 
the absence of an applied magnetic field.  Such a scenario is predicted 
for a variety of different systems including one dimensional (1D) ballistic 
quantum wires \cite{wang1,Goldgpar},  the two dimensional (2D) electron gas
\cite{2DEGpol}, three dimensional (3D) metal nanowires \cite{zabalaPRL} 
and circular quantum dots \cite{koskinenPRL97}. 
In the case of 1D, interactions become increasingly important at low densities and  
models such as the Tomanaga-Luttinger liquid theory \cite{Luttinger} are required to 
describe them. Despite the large exchange energy present in low-density 
1D systems there are strict theoretical arguments against magnetic 
ordering \cite{leibmattis} and   the notion of a 1D spin-polarised ground state remains the 
subject of wide debate, in particular since the important experimental 
results of Thomas {\it et al.} \cite{Thomasspin1st}.  Here we 
present experimental results taken on semiconductor quantum wires in  zero 
magnetic field that provide 
strong evidence in favor of a spin energy gap developing in the 1D region.  
This density-dependent energy gap between spin-up and spin-down electrons is 
revealed in transport measurements as an anomalous conductance feature in the range $0.5 - 0.7 \times 2e^2/h$.  

A feature near $0.7 \times 2e^2/h$ can be seen in some of the earliest 
1D ballistic transport measurements on quantum point contacts \cite{Wharam,vanWeesQPC}. In 1996 
Thomas {\em et al.} \cite{Thomasspin1st} 
revealed that the anomalous feature was related to spin by 
showing that it evolves smoothly into 
the Zeeman spin-split level at $0.5 \times 2e^2/h$ with an in-plane 
magnetic field.   Since that time experimental studies have concentrated 
on the behavior of the anomaly as a function of temperature, source-drain  
bias, magnetic field, thermopower, wire length and density \cite{Thomasint, BKQWAPL, ct-liang,
kristensenPRB, newthomas, Pyshkin, appleyard, reilly}. Together with these 
investigations, numerous mechanisms to explain the origin of the conductance feature 
have been proposed \cite{schmeltzer, flambaum, spivak, wingreen1, sushkov, bruus2, wingreen2}.  Amongst the most 
compelling of these models is the notion of Fermi-level pinning 
in the presence of a static spin energy gap \cite{bruus2, wingreen2}.
\begin{figure}
\begin{center}
\includegraphics[width=6.5cm]{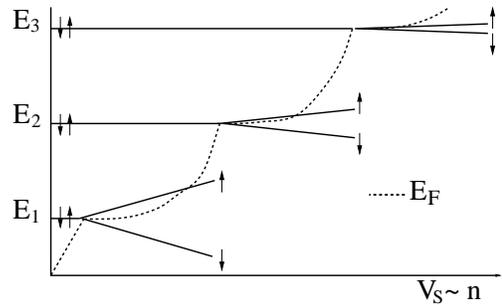}
\caption{Phenomenological picture of a density dependent spin gap opening 
linearly with increasing density ($n$) or gate voltage ($V_S$). $E_1, E_2, E_3$ 
indicate the 1D sub-band edges.  The Fermi level (dashed line) is non-linear 
with density $n$ due to the singularity in the 1D 
density of states.   }
\vspace{-1cm}
\end{center}
\end{figure}
\begin{figure}
\begin{center}
\includegraphics[width=7.0cm]{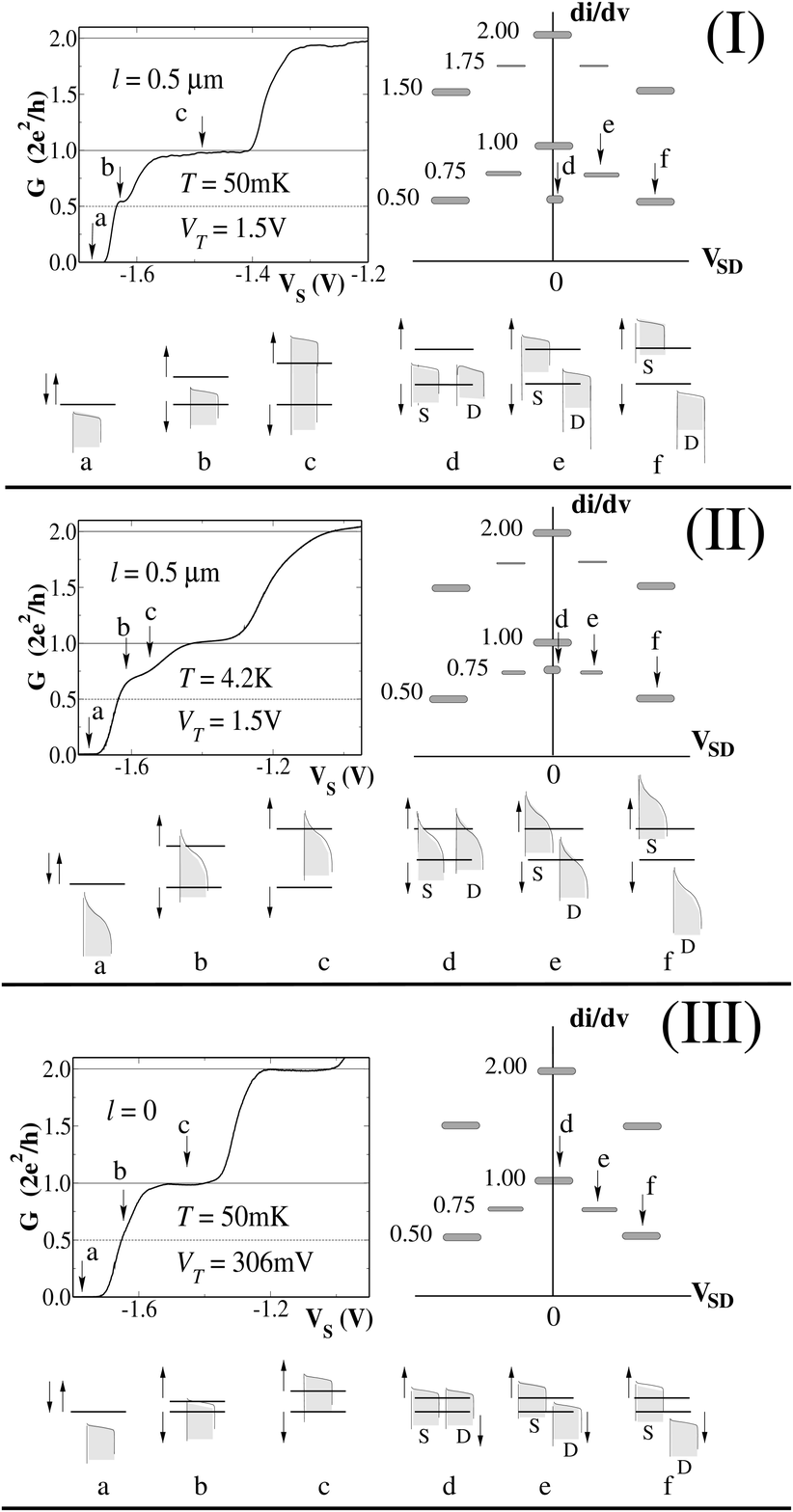}
\caption{Schematic showing the three main scenarios that lead to 
features near $0.5 \times 2e^2/h$ and $0.7 \times 2e^2/h$ in 
the conductance at both zero (left) and finite source - drain bias (right). 
In the lower portion of each graph the horizontal lines indicate the sub-band edges and 
shaded regions represent the Fermi distributions.      Scenario (I) 
occurs if the spin gap is large in comparison to $kT$.   Scenario (II) occurs 
at high temperature when $kT$ is close to the size of the spin gap.  Finally scenario (III) 
occurs for the case of weak spin splitting.}
\vspace{-1cm}
\end{center}
\end{figure}

In this work we propose, and present supportive experimental data for, an alternative 
simple phenomological model that appears to explain the characteristic details of the $0.7$ feature by means of 
a {\em density dependent} spin gap arising in the region of the quantum 
wire.  Key differences 
exist between this simple model and other explanations based on Fermi level 
pinning. As depicted in Fig.(1), the spin levels are degenerate when the 
sub-band is empty, with the spin gap opening up as the carrier density in 
the 1D sub-band is increased.   Consequently this model also explains the 
absence of conductance plateaus at $0.25 \times 2e^2/h$ in the presence of 
a finite source - drain (S-D) bias.  

In contrast to the case of a {\it static} spin polarisation, the sub-bands remain 
spin degenerate until they are populated, after which the spin gap opens 
with increasing density.  This behavior, suggestive of many-body 
interactions, is consistent with a spin polarisation driven by exchange as 
predicted by Wang and  Berggren \cite{wang2}. In their calculations the 
polarisation weakens as higher sub-bands are populated (see Fig.1). The 
non-linear dependence of Fermi energy $E_F$ on density or gate voltage in 
Fig.1 is a consequence of the singularity in the 1D density of states, 
$\rho \sim E^{-1/2}$ \cite{wingreen1}.

Our proposed model is consistent with conductance data we have obtained on GaAs/AlGaAs 
quantum wires free from the disorder associated with modulation 
doping \cite{BKQWAPL}. Although we illustrate our simple model via comparison with this data, the model is not limited to 
these samples but appears to be consistent with the key published results 
\cite{Thomasspin1st,Wharam,vanWeesQPC,Thomasint,BKQWAPL,ct-liang,newthomas,kristensenPRB,Pyshkin,appleyard,reilly}.

Turning now to the experimental data, we note that the presence and 
shape of conductance anomalies depend on the rate at which the spin gap opens with 
1D density and are likely to be sample dependent.  
Figure 2 depicts the three main scenarios that may arise. Scenario (I)  
occurs if spin splitting takes place quickly with increasing density, so that an appreciable energy gap develops 
in comparison to the thermal energy $kT$.  In this case we see a fully 
resolved spin-split plateau near $0.5 \times 2e^2/h$ in linear response 
conductance $G$ and no feature near $0.7 \times 2e^2/h$.  This is shown on the 
left of Figure 2(I), for a quantum wire of length $l = 0.5 \mu m$ at $T$ = 50mK.  
The right side of the figure shows the dependence of the differential 
conductance ($di/dv$) with finite source-drain bias, 
where the thick lines represent conductance plateaus.    Due to an averaging 
of the conductance at the chemical potential of the source $S$ and drain $D$,  
half-plateaus at  0.5 and 1.5  $ \times  2e^2/h$ 
occur at finite bias when the two potentials differ by one
sub-band \cite{patel, moreno}. 

In a large applied magnetic field the Zeeman energy lifts the spin 
degeneracy, such that plateaus at $0.5, 1.5, 2.5 \dots \times 2e^2/h$ are 
resolved in the the conductance \cite{Wharam}. The simultaneous application 
of a finite S - D bias then produces additional quarter plateaus at 
$0.25, 0.75, 1.25, 1.75 \dots \times 2e^2/h$ \cite{patelparrab}. However, 
in the case of a density dependent spin-gap at $B$ = 0, the plateaus at 
$0.25$ and $1.25 \times 2e^2/h$ will be absent, since the 1D electron 
density is not large enough to appreciably open the spin gap.    

Scenario (II) considers the case where the thermal energy $kT$ is 
comparable to the spin gap. In this case no feature near  $0.5 \times 2e^2/h$ will be
resolved as $E_F$ crosses the band edges. Instead, as $E_F$ approaches the upper spin band-edge, the spin-gap 
continues to open so that the number of electrons which thermally 
populate the upper spin band remains approximately constant. A 
quasi-plateau near $0.7 \times 2e^2/h$ therefore occurs due to the 
simultaneous increase of both $E_F$ and the upper spin-split sub-band,  
(shown in the data for the same $l = 0.5 \mu m$ wire at $T$ =
4.2K).   In this model the quasi-plateau can occur anywhere in the range 
$0.5 - 1.0 \times 2e^2/h$, as has been observed experimentally 
\cite{newthomas,reilly}.

The right side of Fig. 2(II) illustrates the behavior of the 
differential conductance at elevated temperatures.  In contrast to the low 
temperature case of scenario (I), the feature remains close to 
$0.75 \times 2e^2/h$ even when the S - D bias is close to zero 
(Fig. 2(II) d \& e).

Scenario (III) illustrates the case for where the spin-splitting is weak 
and only grows slowly with increasing density.  At 
low temperatures a feature near $0.5 \times 2e^2/h$ is absent if the spin gap 
remains small in comparison to $kT$. We illustrate this scenario with 
data taken on a $l$ = 0 quantum wire at $T$ = 50mK.  Although there is no 
evidence for a spin gap at zero S - D bias, the spin gap can still be 
observed as a feature at $0.75 \times 2e^2/h$
in the differential conductance as shown in the schematic on the right.  
This is because the splitting is too small to be resolved at low densities
where $G < 2e^2/h$ at $V_{SD}$ = 0. However, a moderate 
S - D bias evolves the $1.0 \times 2e^2/h$ plateau into a 
feature near $0.75 \times 2e^2/h$ as the source and drain potentials differ 
by one spin sub-band (Fig.  2(II)e).   Such behavior has been observed by Kristensen {\it 
et al.} \cite{kristensenPRB,foot2}.

\begin{figure}
\begin{center}
\includegraphics[width=7.0cm]{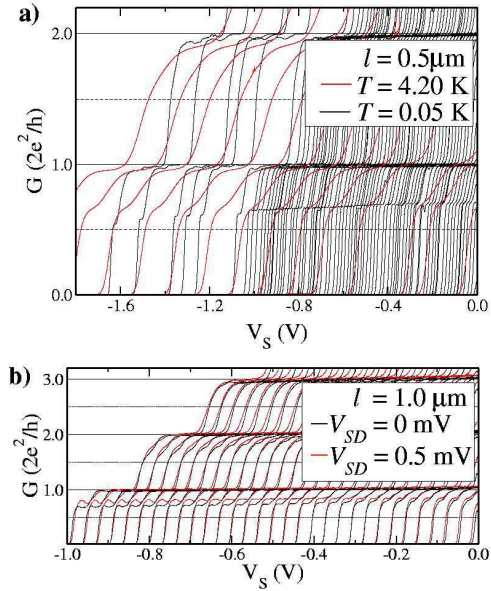}
\caption{{\bf (a)} Conductance of a $l=0.5 \mu m$ quantum wire as a function of
side gate voltage $V_S$ for top gate voltages in the range; $V_T$ = 420mV- 1104mV 
(right to left). {\bf (b)} Conductance of a $l=1.0 \mu m$ quantum 
wire as a function of $V_S$ for $V_T$ in the range; 270mV - 800mV (right to left). $T$ = 50mK. 
}
\vspace{-1cm}
\end{center}
\end{figure}
\begin{figure}
\begin{center}
\includegraphics[width=7.0cm]{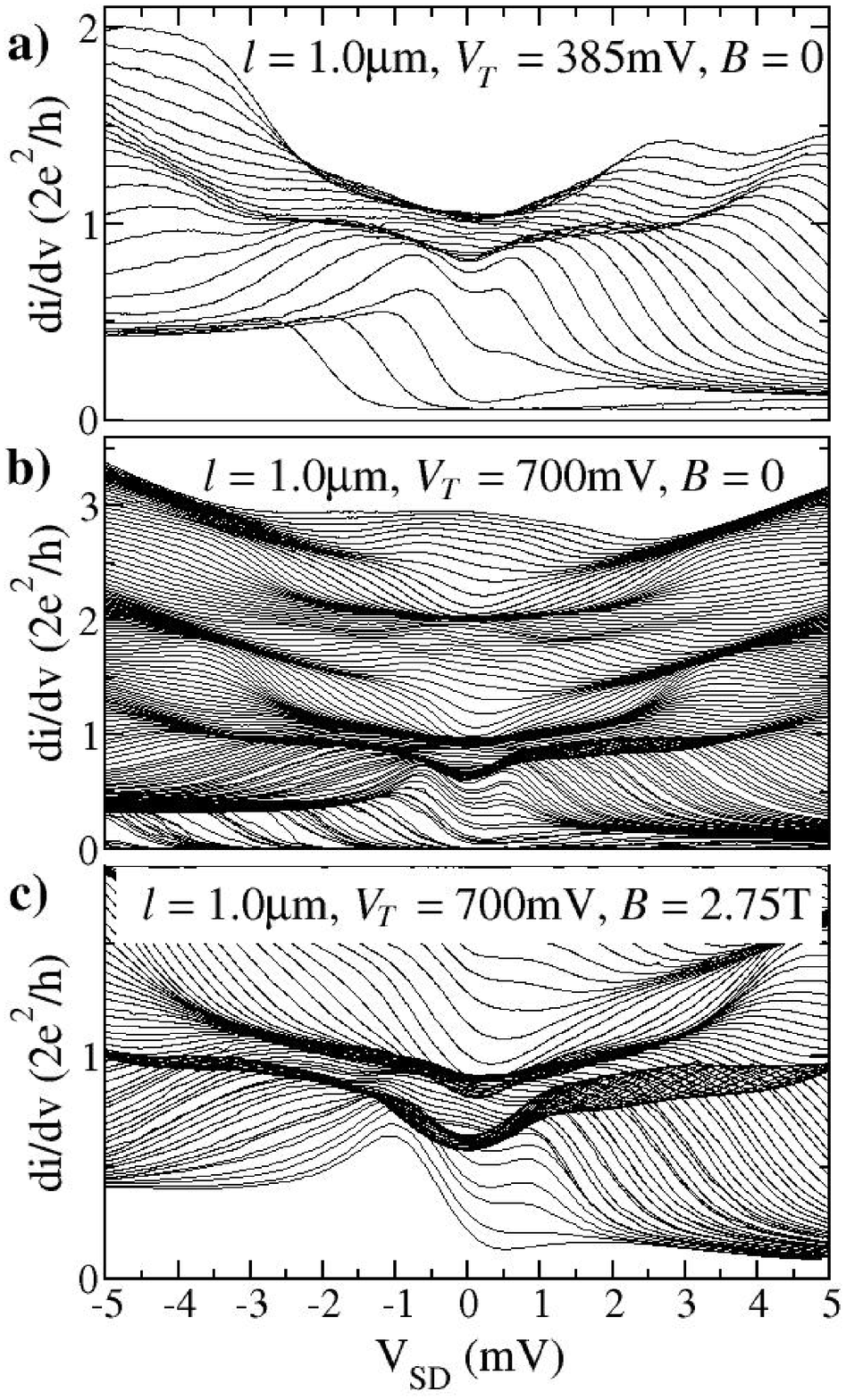}
\caption{Differential conductance of a $l=1.0\mu m$
quantum wire.  {\bf (a)} $V_S$ = 0 to -1200mV in -5mV steps. {\bf (b)} $V_S$ = -850mV to
-1400mV in -0.5mV steps.  {\bf (c)} $V_S$ = -900mV to -1400mV in -1mV steps. All 
data taken at $T$ = 50mK.  }
\vspace{-1cm}
\end{center}
\end{figure}
We now present conductance data taken on two different samples which 
support our model. The devices are fabricated from ultra low-disorder GaAs/AlGaAs heterostructures 
with electron mobilities in the range $4 - 6 \times 10^6 cm^2 V^{-1}s^{-1}$.  
Using electron beam lithography the top layer of the heterostructure is sectioned into 
three separately controllable gates with the  middle or top gate  being  biased 
positively ($V_T$) to control both the density  in the 2D reservoirs ($ n_{2D} = 1.5 - 7.5 \times 10^{11}$cm$^{-2}$) and in 
the 1D channel \cite{BKQWAPL}.
The side gates are negatively biased ($V_{S}$) and simultaneously control the 1D 
density and the transverse potential.  Careful control of all 
three  independent gates allows the $0.7 \times 2e^2/h$ feature to be studied as a 
function of both density and potential profile.  We find that the feature tends towards 
$0.5 \times 2e^2/h$ with {\em increasing} top gate bias and length of the 1D 
region \cite{reilly}.   

Figure 3(a) shows data taken on a quantum wire of length 
$l = 0.5 \mu m$ at $T$ = 4.2K and $T$ = 50mK. The data 
taken at $T$ = 50mK show a characteristic evolution towards fully 
resolved spin-splitting with the $0.7$ feature moving 
closer to $0.5 \times 2e^2/h$ with increasing top gate bias (right to left). 
In the context of the model this evolution is consistent with 
the spin gap opening more rapidly with 1D density $n$, for larger $V_T$, 
(i.e. moving from scenario III to scenario I with increasing $V_T$) 
although the exact mechanism behind the dependence on $V_T$ is 
presently unknown \cite{foot}. In particular, the left-most traces from 
Fig. 3(a) are consistent with scenario (I) at $T$ = 50mK and scenario (II) 
at $T$ = 4.2K, where the feature has risen from $0.55$ to $0.7 \times 2e^2/h$.  

At $T$ = 4.2K the position of the feature 
does not evolve with top gate bias (2D density) but remains close to $0.65 \times 2e^2/h$ 
at these higher temperatures, as both of the 1D spin bands are populated 
({\it c.f.} Fig 2.(II)), and the position of the feature is rather 
insensitive to small changes in the spin-gap.  

The shape of the feature will depend on the slope of the Fermi function as 
it crosses the second spin-band edge: at high temperatures the broad Fermi 
function produces a broad quasi-plateau.  At low temperatures, depending 
on the size of the spin gap, the sharp Fermi function will either produce a 
small but sharp feature at $0.5 \times 2e^2/h$ (strong splitting, 
Scenario(I)) or a weak inflection (weak splitting, Scenario(II)) in the conductance.  

Figure 3(b) explores the effect of a constant dc offset bias on 
both the shape and position of the conductance feature. At 
$V_{SD}$ = 0 (black curves) we observe evolution of the feature from $0.75$ 
towards $0.5 \times 2e^2/h$ with increasing $V_T$, consistent with the low 
temperature data for the $l = 0.5 \mu m$ quantum wire shown in Fig.(3a)
The application of a DC offset bias $V_{SD}$ = 0.5mV causes the feature to 
rise towards $0.75 \times 2e^2/h$, with a much weaker dependence on $V_T$.   
 This behavior mirrors that in Fig 3(a) since increasing $V_{SD}$ or 
increasing $T$ will distribute electrons between both spin bands.  

With the application of a dc source - drain  bias the spin gap can be studied as a 
function 1D density, controlled by the side gate bias $V_S$.  Figure 4 shows the 
differential conductance ($di/dv$) of a $l=1 \mu m$ 
quantum wire as a function of $V_{SD}$ for different side gate 
voltages at $T$ = 50mK. We compare the differential conductance at 
two different 2D densities ($V_T$ = 385mV, 
$n_{2D} = 2.6 \times 10^{11}cm^{-2}$ in (a) and  
$V_T$ = 700mV, $n_{2D} = 5.4 \times 10^{11}cm^{-2}$   in (b)).  In these 
plots conductance plateaus appear as a grouping of individual curves, as 
can clearly be seen in Fig.  4(b) where plateaus occur at $2$, $1$ and 
$0.75 \times 2e^2/h$ for $V_{SD}$ = 0. In both Figs. 4(a) and (b)
clear half-plateaus at $0.5$ \& $1.5 \times 2e^2/h$ can be seen 
developing near $V_{SD} = \pm 1.5 mV$, (although the $0.5$ half-plateau on the right 
side of each graph is suppressed due to the asymmetric bias across the 
constriction near pinch-off \cite{moreno}).

Most importantly there are no plateaus near $0.25$ and $1.25 \times 2e^2/h$, 
despite the presence of strong features at $0.75 - 0.9 \times 2e^2/h$.  
This cannot be explained by a static spin gap, but is a natural consequence 
of a density dependent gap: at low 1D densities (small conductances) the 
spin gap has not yet developed, but at larger densities the gap opens up 
and features are observed at $\approx 0.75 \times 2e^2/h$.  

Further evidence for the density dependence of the gap can be seen
in the region close to zero bias where a characteristic `cusp' 
feature is seen below the $1 \times 2e^2/h$ plateau. Traces associated with 
this feature start at $0.5 \times 2e^2/h$ at zero source - drain bias 
(scenario I) and move towards $0.75 \times 2e^2/h$ as $V_{SD}$ is 
increased. As the 1D density is increased, (by altering $V_S$) the spin-gap 
widens, and a larger $V_{SD}$ must be applied before the conductance 
increases above $\approx 0.5 \times 2e^2/h$.  The cusp feature is a result 
of many of these traces overlapping, and the strength and width of the 
feature is a measure of how large the spin gap is, and how rapidly it changes 
with 1D density.  The increasing strength of the spin splitting with 
increasing $V_T$  can be seen in Figs. 4(a) and (b), where the cusp 
becomes deeper and more strongly resolved.   With the application of an 
external parallel magnetic field the spin gap opens further and the cusp 
widens (Fig.4(c)) \cite{contacts}. In future it may be possible to analyse 
the detailed shape of the cusp feature in order to obtain quantitative 
information about how the spin gap evolves with 1D density and parallel 
magnetic field.  

In conclusion, we have presented a simple model to explain the $0.7 \times 2e^2/h$ conductance feature 
in terms of a density dependent spin polarisation arising in the 1D region. 
While our phenomenological model is consistent both with  experimental data 
for ultra low-disorder quantum wires presented here, and with other 
published data, a detailed microscopic explanation of the 
spin polarisation is still lacking.  In particular, 
the  rate at which the spin-splitting grows with 1D density is sample dependent and seems to depend on
 the length of the 1D region, the  surface gate geometry, and the 2D electron density. 
A detailed understanding of the spin polarisation 
will have implications for 1D electron transport in 
mesoscopic devices and may have important applications to the field of spintronics.  

We thank D. Barber, R.  P.   Starrett, and N.  E.  Lumpkin for technical 
support. This work was funded by the Australian Research Council.  


\begin{thebibliography}{10}
\small
\bibitem[*]{email}
email: djr@jupiter.phys.unsw.edu.au

\bibitem[$\dagger$]{bruce}
Now at The Laboratory for Physical Sciences, the University of Maryland, 
College Park, MD.  

\bibitem{wang1}
C.   K.   Wang and K.  F.   Berggren, Phys.   Rev.   B.   {\bf 54},  R14257 (1996).

\bibitem{Goldgpar}
A.   Gold and L.   Calmels, Phil.   Mag.   Lett.   {\bf 74},  33  (1996).


\bibitem{2DEGpol}
D. Varsano, S.Moroni and G. Senatore EuroPhys. Lett. {\bf 53} (3) 348 (2001).


\bibitem{zabalaPRL}
N.    Zabala, M.    J.    Puska and R.     M.    Nieminen, Phys.    Rev.    
Lett.    {\bf 80}, 3336 (1998).


\bibitem{koskinenPRL97}
M. Koskinen, M. Manninen and S. M. Reimann, Phys. Rev. Lett.  {\bf 79}, 1389 (1997).

\bibitem{Luttinger}
J.~M. Luttinger, J. Math. Phys. {\bf 4},  1154  (1963).

S.   Tomonaga, Prog.   Theor.   Phys.  , {\bf 5}(4), 544 (1950).


\bibitem{leibmattis}
E. Lieb and D. Mattis, Phys Rev.   {\bf 125}, 164 (1962).

\bibitem{Thomasspin1st}
K.~J. Thomas {\it et~al.}, Phys. Rev. Lett. {\bf 77},  135  (1996).

\bibitem{Wharam}
D.   A.  Wharam {\it et al.}, J.   Phys.  C {\bf 21}, L209 (1988).

\bibitem{vanWeesQPC}
B.  J.   {van Wees} {\it et~al.  }, Phys.   Rev.   Lett.   {\bf 60},  848  (1988).

\bibitem{Thomasint}
K.~J. Thomas {\it et~al.}, Phys. Rev. B {\bf 58},  4846  (1998).

\bibitem{BKQWAPL}
B.  ~E.   Kane {\it et~al.  }, Appl.   Phys.   Lett.   {\bf 72},  3506  (1998).

\bibitem{ct-liang}
C. -T. Liang  {\it et al.    }, Phys.     Rev.    B  {\bf 60}, 10687 (1999).

\bibitem{newthomas}
K.   J.   Thomas   {\it et~al.  }, Phys.   Rev.   B {\bf 61}, R13365 (2000).

\bibitem{kristensenPRB}
A.   Kristensen {\it et al.} Phys.   Rev. B {\bf 62}, 10950 (2000).  

\bibitem{Pyshkin}
K.   S.   Pyshkin  {\it et al.  } Phys.   Rev.   B {\bf 62} 15842 (2000).

\bibitem{appleyard}
N.  J.  Appleyard  {\it et al.    }, Phys.     Rev.    B {\bf 62},  R16275  (2000).

\bibitem{reilly}
D. J. Reilly {\it et al.  }, Phys.   Rev.  B {\bf 63},  R121311  (2001).

\bibitem{spivak}
B. Spivak and F. Zhou, Phys.   Rev.   B {\bf 61}, 16730 (2000).

\bibitem{sushkov}
O. P. Sushkov, Phys.  Rev.  B{\bf 64}, 155319 (2001). 

\bibitem{flambaum}
V.~V. Flambaum and M.~Yu. Kuchiev, Phys. Rev. B {\bf 61}, R7869 (2000).

\bibitem{schmeltzer} 
D. Schmeltzer  {\it et~al.  } Phil. Mag. B {\bf 77} 1189 (1998).

\bibitem{wingreen1}
K. Hirose and N.  S.  Wingreen Phys. Rev. B {\bf 64}, 073305 (2001).

\bibitem{wingreen2}
K. Hirose, S. S. Li and N.  S.  Wingreen Phys. Rev.  B {\bf 63}, 033315 (2001).   

\bibitem{bruus2}
H. Bruus {\it et al.  } Physica E 2000.

\bibitem{wang2}
C. K. Wang and K.  F.   Berggren, Phys.   Rev.  B {\bf 57}, 4552 (1998).


\bibitem{patel}
N. ~K. Patel  {\it et~al.},  Phys. Rev. B. {\bf 44}, 13549 (1991).

\bibitem{moreno}
L. Martin-Moreno {\it et al.  } J.   Phys.  C {\bf 4} 1323 (1992).

\bibitem{patelparrab}
N. K. Patel {\it et al.   } Phys.   Rev.   B {\bf 44} 10973 (1991).

\bibitem{foot2}
We note that the spin gap should remain constant
with the application of a source - drain bias since, to a first 
approximation, the bias does not change the 1D density.        


\bibitem{foot}
Although the 2D density grows linearly with $V_T$, for a fixed conductance the 1D saddle potential 
($\omega_x/\omega_y$) and the 1D density vary only a small amount as $V_T$ 
is increased. 

\bibitem{contacts}
Unfortunately qualitative comparison is made difficult by the 
effect of the magnetic field on the ohmic contact resistance.

\end{thebibliography}
\end{document}